\documentclass[12pt,a4paper]{article}

\usepackage{graphicx}

\usepackage{amssymb,amsmath,amsfonts}
\usepackage{axodraw,epsfig}

\def\,{\ifmmode\mskip\thinmuskip\else\leavevmode\thinspace\fi}

\begin{document}

\title{Polarization transfer measurements of proton form factors: deformation by initial
collinear photons}
\date{18. July 2005} % optional
\author{S.Dubni\v{c}ka$^1$, E.Kuraev$^2$, M.Se\v{c}ansk\'y$^{2}$ and A.Vinnikov$^2$}
\maketitle
\begin{center}{
$^{1}$ \it Institute of Physics, Slovak Academy of Sciences,
D\'ubravsk\'a 9, 84511 Bratislava, Slovak Republic \\
$^{2}$ \it Bogoliubov Lab. of Theor. Physics, JINR Dubna, 141980
Dubna, Russia \\}
\end{center}

\vspace*{2cm}
\begin{abstract}
It is demonstrated that an emission of collinear photons by the
polarized initial electron in elastic electron-proton polarization
transfer scattering leads to an apparent shifting of real events
with small momentum transfer into the data sample with large
momentum transfer. Effectively this shows a fictive enhancement of
the cross section at large momentum transfer. However, the
enhancement is different for transverse and longitudinal
polarizations of the recoil proton. The former is responsible for
a deformation of results when extracting the proton
electromagnetic form factors ratio from the data on
electron-proton polarization transfer scattering. Nevertheless,
this effect does not explain the suppression of the Dirac form
factor at large momentum transfer completely.
\end{abstract}

In the past few years attention to the proton elastic form
factors, which have always played an essential role in
understanding of the nucleon electromagnetic structure, was
reinforced. The reason for that was an appearance of JLab proton
polarization data \cite{clas1,clas2,clas3} on the ratio
$G_{Ep}(Q^2)/G_{Mp}(Q^2)$ measured by the method of polarization
transfer \cite{ahiezer,arnold}. In contrast to the apparently
well-established experimentally by the so-called Rosenbluth
technique \cite{rosen} ratio $\mu_p G_{Ep}(Q^2)/G_{Mp}(Q^2)
\approx 1$, which in a wide region of $Q^2$ (0.1 GeV$^2 < Q^2 < $
33.4 GeV$^2$) \linebreak is approaching one, the new results in
the range 0.49 GeV$^2 \leq Q^2 \leq $5.54 GeV$^2$ reveal rapid
fall of the ratio as $Q^2$ increases.

The inconsistency in the results obtained by the above mentioned
two different methods produced a broad discussion on the
reliability of the methods and the accuracy of the one photon
exchange approximation (see Refs.~\cite{arrington} and references
therein) as well. As to the latter, it is known that in scattering
of electrons and positrons on protons the contributions from two
photon exchange diagram are small \cite{mar,meren}. Nevertheless,
new results of two-photon exchange calculations \cite{brodsky}
relying on the knowledge of the proton structure, which contain
more detailed information than the form factors under
consideration, demonstrate the situation to be more complicated.
As a result there is a proposal  \cite{twophotexp} to
reinvestigate the contribution of two photons exchange by a
measurement of the difference of electron and positron cross
sections on the proton more carefully.

The other source of corrections which can influence the extraction
of the form factors from the polarization data is a radiation of
photons by the initial electrons along the beam line. Such photons
are not therefore registered. They take away a part of the
electron energy, then the genuine momentum transfer $Q_p^2$ to the
proton is less than the value obtained from the elastic electron
scattering angle $Q^2=4E_1 E_2 \sin^2\frac{\theta}{2}$, where the
energy $E_2$ of the final electron is not measured and it is
determined from the elastic scattering formula $E_2=ME_1 /\left (
M+2E_1 \sin^2\frac{\theta}{2} \right )$. The cross section of the
process $ep\to e'p'$ (no emission of extra photons) falls rapidly
as $Q^2$ increases. Therefore, if the energy of the final electron
is not measured in order to check the elasticity of the event, the
contribution of the inelastic events to the cross section can be
significant since for them $Q_p^2< Q^2$ even though emission of an
additional photon is suppressed by a factor of $\alpha_{em}$. Such
corrections to the one-photon exchange can be calculated without
model dependent assumptions. This task has been done in many works
(see e.g. the original paper \cite{motsai}) and the corrections
were found to be small \cite{meren}. However, the corresponding
calculations are quite complicated technically and no simple
formula can be presented for their understanding.

In this letter we present a simple estimation of the corrections
arising from the photon radiation by the initial electron.

\begin{figure}
\centering
\begin{minipage}[c]{0.45\hsize}
\begin{picture}(200,100)(0,0)
\ArrowLine(5,95)(100,75)
\ArrowLine(100,75)(195,95)
\Photon(100,75)(100,25){5}{3}
\DashLine(100,75)(157,63){4}
\CArc(100,75)(30,-13,13)
\SetWidth{2.0}
\ArrowLine(5,5)(100,25)
\ArrowLine(100,25)(195,5)
\Text(15,85)[]{$k_1$}
\Text(185,85)[]{$k_2$}
\Text(15,15)[]{$p_1$}
\Text(185,15)[]{$p_2$}
\Text(140,75)[]{$\theta$}
\Text(85,50)[]{$Q^2$}
\end{picture}
\caption{Kinematics of $ep\to e'p'$ reaction. No extra photon emission is assumed.}
\label{kinemat}
\end{minipage}
\hspace*{5mm}
\begin{minipage}[c]{0.45\hsize}
\begin{picture}(200,100)(0,0)
\ArrowLine(5,95)(100,75)
\ArrowLine(100,75)(195,95)
\Photon(100,75)(100,25){5}{3}
\Photon(20,91.84)(50,95){3}{3}
\DashLine(100,75)(157,63){4}
\CArc(100,75)(30,-13,13)
\SetWidth{1.0}
\DashLine(60,95)(60,5){4}
\SetWidth{2.0}
\ArrowLine(5,5)(100,25)
\ArrowLine(100,25)(195,5)
\Text(15,85)[]{$k_1$}
\Text(185,85)[]{$k_2$}
\Text(15,15)[]{$p_1$}
\Text(185,15)[]{$p_2$}
\Text(75,70)[]{$xk_1$}
\Text(75,30)[]{$p_1$}
\Text(140,75)[]{$\theta$}
\Text(115,50)[]{$Q_p^2$}
\end{picture}
\caption{Kinematics of $ep\to e'p'$ reaction. Initial electron emits collinear photons.}
\label{emis}
\end{minipage}
\end{figure}

Before, however, we briefly remind the formulae related to the
method of polarization transfer assuming no emission of extra
photons \cite{ahiezer,arnold}. The longitudinally polarized
electron beam, with energy in lab. frame\footnote{Throughout the
letter, all non-covariant variables are written in the lab.
frame.}
 $E_1$ and polarization degree $\lambda$, scatters
on unpolarized proton target (Fig.~\ref{kinemat}). In the final
state, the knocked out proton is detected and its polarization is
measured. The degree of transverse proton polarization is denoted
by ${\cal P}_x$, the degree of its longitudinal polarization is
denoted by ${\cal P}_z$ and both depend on the electron scattering
angle $\theta$ as follows \cite{ahiezer,arnold}:
\begin{equation}
{\cal P}_x \frac{d\sigma}{d\Omega}=-\lambda \frac{\alpha^2}{Q^2}
\left ( \frac{M}{M+2E_1 \sin^2 \frac{\theta}{2}} \right )^2
\frac{Q}{\sqrt{Q^2+4M^2}}
{\rm ctg} \frac{\theta}{2}~G_{Ep}(Q^2)G_{Mp}(Q^2) ,
\label{polx}
\end{equation}
\begin{equation}
{\cal P}_z \frac{d\sigma}{d\Omega}=-\lambda \frac{\alpha^2}{2M^2}
\left ( \frac{M}{M+2E_1 \sin^2 \frac{\theta}{2}} \right )^2
\sqrt{1+\frac{4M^2}{Q^2+4M^2}{\rm ctg}^2\frac{\theta}{2}}~G_{Mp}^2(Q^2),
\label{polz}
\end{equation}
where $Q^2=-(k_1-k_2)^2$, $M$ is the proton mass. The scattering
angle $\theta$ is related to $Q^2$ by means of the expression
\begin{equation}
\sin^2\frac{\theta}{2}=\frac{Q^2}{4E_1(E_1-\frac{Q^2}{2M})}.
\end{equation}
From Eqs. (\ref{polx}) and (\ref{polz}) the relation
\begin{equation}
R(Q^{2})=\frac{G_{Ep}(Q^2)}{G_{Mp}(Q^2)}= \frac{{\cal
P}_x\frac{d\sigma}{d\Omega}}{{\cal P}_z\frac{d\sigma}{d\Omega}}
\frac{Q^2}{2M^2}{\rm tg}\frac{\theta}{2}
\sqrt{1+\frac{4M^2}{Q^2\sin^2\frac{\theta}{2}}}. \label{ratio}
\end{equation}
follows, which is commonly used to extract the value of
$G_{Ep}(Q^2)/G_{Mp}(Q^2)$ from the polarization transfer data.

Let us now take into account emission of extra photons by the
initial electron\footnote{Photon emission by the final electrons
is not important since the photon in that case should hit the same
detector cell as the final electron. Thus in such case the entire
energy of the electron after collision with the proton is
detected.} (Fig.~\ref{emis}). The process can be split into two
parts: the extra photon emission, described by ${\cal D}$-function
\cite{kuraevdfunc} and the scattering of the slowed down electron
on the proton described by Eqs.~(\ref{polx}),(\ref{polz}). Since
the extra photon is almost collinear to the initial electron, we
take $k_1'=xk_1$ and it follows that the energy of the final
electron $E_2'$ and the momentum transferred to the proton
$Q_p^2=-(p_1-p_2)^2$ are given by
\begin{equation}
E_2'=\frac{xME_1}{M+2xE_1\sin^2\frac{\theta}{2}},
\end{equation}
\begin{equation}
Q_p^2=\frac{4Mx^2
E_1^2\sin^2\frac{\theta}{2}}{M+2xE_1\sin^2\frac{\theta}{2}} < Q^2
= \frac{4M
E_1^2\sin^2\frac{\theta}{2}}{M+2E_1\sin^2\frac{\theta}{2}} .
\label{qeq}
\end{equation}
Then the radiatively corrected expressions for the cross sections (\ref{polx}),(\ref{polz})
read
\begin{equation}
\left ( {\cal P}_x \frac{d\sigma}{d\Omega} \right )_{corr}
=-\lambda \int\limits_{x_0}^{1} dx {\cal D}(x)
\frac{\alpha^2}{Q_p^2} \left ( \frac{M}{M+2xE_1 \sin^2
\frac{\theta}{2}} \right )^2 \frac{\sqrt{Q_p^2}~{\rm ctg}
\frac{\theta}{2}}{\sqrt{Q_p^2+4M^2}}
~R(Q_p^2)_{corr}G^2_{Mp}(Q_p^2)_{corr} , \label{polxpr}
\end{equation}
\begin{equation}
\left ( {\cal P}_z \frac{d\sigma}{d\Omega} \right )_{corr}
=-\lambda \int\limits_{x_0}^{1} dx {\cal D}(x)
\frac{\alpha^2}{2M^2} \left ( \frac{M}{M+2xE_1 \sin^2
\frac{\theta}{2}} \right )^2 \sqrt{1+\frac{4M^2{\rm
ctg}^2\frac{\theta}{2}}{Q_p^2+4M^2}} ~G_{Mp}^2(Q_p^2)_{corr},
\label{polzpr}
\end{equation}
where the ${\cal D}$-function is given by \cite{kuraevdfunc}
\begin{eqnarray}
{\cal D}(x)=\frac{\beta}{2}\left [ \left ( 1+\frac{3}{8}\beta \right )
(1-x)^{\frac{\beta}{2}-1} - \frac{1}{2}(1+x) \right ] - \nonumber \\
\frac{\beta^2}{32}\left [ 4(1+x)\ln(1-x) + \frac{1+3x^2}{1-x}
\ln x + 5 + x \right ] + {\cal O}(\beta^3),
\end{eqnarray}
\begin{equation}
\beta=\frac{2\alpha}{\pi}\left [ \ln \left ( \frac{Q_p^2}{m_e^2} \right ) -1
\right ]
\end{equation}
and $x_0$ is determined by the minimal energy $E_{2min}$
the final electron has to carry to be detected:
\begin{equation}
x_0=\frac{ME_{2min}}{ME_1-2E_1E_{2min}\sin^2\frac{\theta}{2}}\sim\frac{E_{2min}}{E_1}.
\end{equation}

As it can be seen from Eq.~(\ref{qeq}), at small $x$ and fixed
$\theta$ the integrand in Eq.~(\ref{polxpr}) behaves like $1/x$,
while the integrand of Eq.~(\ref{polzpr}) behaves like $const(x)$.
Therefore, extracted from the polarization data ratio of electric
and magnetic form factors is sensitive to the value of $x_0$. The
other very important issue is that the radiative correction is
responsible for the effect of relative enhancement of the
differential cross section at large $Q^2$ with regard to the
radiatively non corrected result
\cite{motsai,kuraevdfunc,resonance}. Indeed, as the value of $Q^2$
rises, the cross sections (\ref{polx}),(\ref{polz}) fall rapidly.
However, the expressions for the radiatively corrected cross
sections (\ref{polxpr}),(\ref{polzpr}) contain integration over
the momentum transferred to the proton. At small $x_0$, the main
contribution to the integrals (\ref{polxpr}),(\ref{polzpr}) comes
from the region of small $Q_p^2$. Thus, even though the
corrections are suppressed by the electromagnetic coupling
constant, they become relatively large as the value of $Q^2$
rises.

To verify how large numerically these effects are, we have
computed the corrected ratio (see Fig.3)
\begin{equation}
R(Q^2)_{corr}=\frac {\left ( {\cal
P}_x\frac{d\sigma}{d\Omega}\right )_{corr}} {\left ( {\cal
P}_z\frac{d\sigma}{d\Omega}\right )_{corr}} \left(
\frac{Q^2}{2M^2}{\rm tg}\frac{\theta}{2}
\sqrt{1+\frac{4M^2}{Q^2\sin^2\frac{\theta}{2}}}\right).
\end{equation}
at different values of $x_0$.
%The first two ratios represent the relative value of radiative
%corrections at different values of $Q^2$. They are shown in
%Figs.~\ref{res1},\ref{res2} for different values of $x_0$.

For the sake of simplicity we have taken
$G_{Ep}=1/(1+Q_p^2/0.71)^2$ and $G_{Mp}=\mu_p/(1+Q_p^2/0.71)^2$,
i.e. $\mu_pG_{Ep}(Q^2)/G_{Mp}(Q^2)=1$. As it can be seen, the
radiative corrections are very large if $x_0$ is small. Therefore,
extraction of the form factors requires a careful measurement of
energy of the final electron. To make the extraction directly, the
elastic scattering formula for $E_2$ has to be checked carefully,
that provides a firm cut off for the collinear photons emission.
\begin{figure}[t]
\begin{center}
\includegraphics[width=0.6\textwidth]{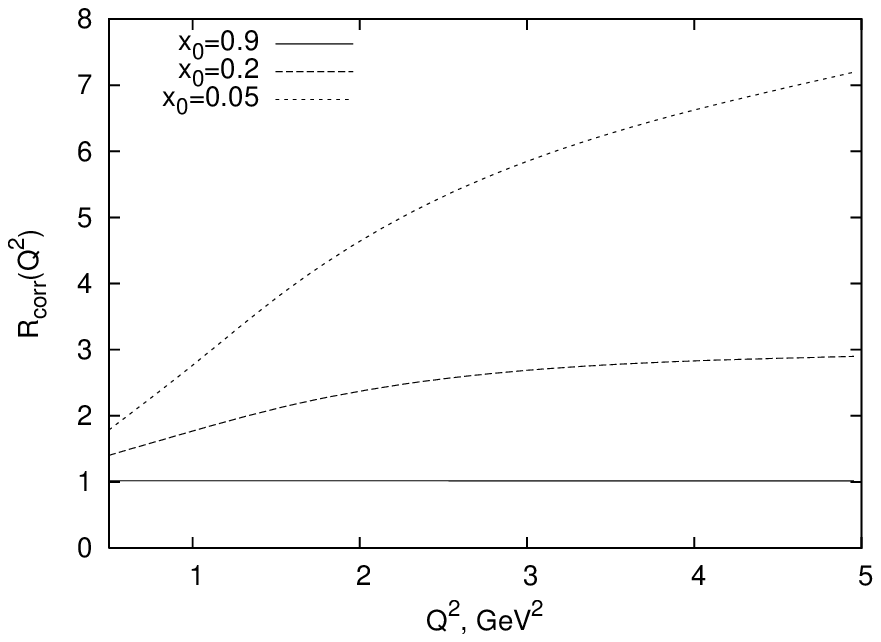}
\caption{Ratios $R(Q^2)_{corr}$ for different
$x_{0}$}\label{result}
\end{center}
\end{figure}
At the experiment, in which such cut off is used to indemnify
elastic kinematics, the radiation of photons by the initial
electrons along the beam line is suppressed. In this case
two-photon exchange contributions can play an important role.
However, their complete evaluation can not be carried out in a
model-independent way \cite{brodsky} and one is forced to restrict
himself to evaluation of box-diagrams. Here, one can estimate in a
model-independent way diagrams with one-proton and also
delta-resonance in the intermediate state. At the unpolarized
process such contributions do not contain terms with $Ln
\frac{Q^{2}}{m^{2}_e}$ and they are finite for $m_e\rightarrow 0$.

Investigations of such contributions are in progress.

\section*{Acknowledgements}
This work is supported by RFBR grants 04-02-16445, 03-02-17291,
and also the aimed BLTP-IP SAS project, as well as the VEGA Grant
2/4099/24 of the Slovak Grant Agency for Sciences. A.V.
acknowledges support of Alexandr von Humboldt foundation. One of
us (E.A.K.) is grateful to D.Toporkov for discussion of
experimental set-ups.

\end{document}